\documentclass[10pt,A4paper,amsmath,amssymb,aps,etoolbox,floatfix,letterpaper,
noeprint,nofootinbib,
noshowpacs, 
preprintnumbers,reprint,showkeys,
twocolumn]
{revtex4}

\newcommand{\beq}{\begin{equation}}

\newcommand{\eeq}{\end{equation}}
\newcommand{\ds}{\displaystyle}

\usepackage{array}
\usepackage{bm}       
\usepackage{color}
\usepackage{dcolumn}  
\usepackage{epsfig}
\usepackage{float}
\usepackage{gensymb}
\usepackage{graphicx} 
\usepackage{hyperref}
\usepackage{longtable}
\usepackage{makecell}
\usepackage{mathtools}
\usepackage{natbib}
\usepackage{pdflscape}
\usepackage{sidecap}
\usepackage{slashbox}
\usepackage[normalem]{ulem}
\usepackage{url}

\graphicspath{%
    {converted_graphics/}
    {/}
}
\begin{document}

\title[Cosmology and SME]{Cosmology and the massive photon frequency shift in the Standard-Model Extension}

\author
{
Alessandro D.A.M. Spallicci\textsuperscript{a,b,c,d}\footnote{email: spallicci@cnrs-orleans.fr
}}
\author{Jos\'e A. Helay\"el-Neto\textsuperscript{e}\footnote{email: josehelayel@gmail.com 
}}
\author{Mart\'in L\'opez-Corredoira\textsuperscript{f,g}\footnote{email: fuego.templado@gmail.com
}}
\author{Salvatore Capozziello\textsuperscript{h,i}\footnote{email: salvatore.capozziello@unina.it\\
}}

\affiliation{
\mbox{\textsuperscript{a}Universit\'e d'Orl\'eans, Observatoire des Sciences de l'Univers en r\'egion Centre (OSUC) UMS 3116} \\
\mbox{1A rue de la F\'{e}rollerie, 45071 Orl\'{e}ans, France}
\vskip1pt
\mbox{\textsuperscript{b}Universit\'e d'Orl\'eans, Collegium Sciences et Techniques (CoST), P\^ole de Physique}\\
\mbox{Rue de Chartres, 45100 Orl\'{e}ans, France} 
\vskip1pt
\mbox{\textsuperscript{c}Centre National de la Recherche Scientifique (CNRS)}\\
\mbox{Laboratoire de Physique et Chimie de l'Environnement et de l'Espace (LPC2E) UMR 7328}\\
\mbox {Campus CNRS, 3A Avenue de la Recherche Scientifique, 45071 Orl\'eans, France}
\vskip1pt
\mbox{\textsuperscript{d}Universidade do Estado do Rio de Janeiro (UERJ), Instituto de F\'{i}sica, Departamento de F\'{i}sica Te\'{o}rica}\\
\mbox{Rua S\~ao Francisco Xavier 524, Maracan\~a, Rio de Janeiro, 20550-013, Brasil}
\vskip1pt
\mbox{\textsuperscript{e}Centro Brasileiro de Pesquisas F\'{\i}sicas (CBPF)}\\
\mbox{Departamento de Astrof\'{\i}sica, Cosmologia e Intera\c{c}\~{o}es Fundamentais (COSMO)}\\
\mbox {Rua Xavier Sigaud 150, 22290-180 Urca, Rio de Janeiro, RJ, Brasil}
\vskip1pt
\mbox{\textsuperscript{f}Instituto de Astrof{\'i}sica de Canarias (IAC)}\\
\mbox {/ V{\' i}a L{\'a}ctea, s/n, 38205 La Laguna (Tenerife), Espa\~na}
\vskip1pt
\mbox{\textsuperscript{g}Universidad de La Laguna, Facultad de Ciencias, Secci{\' o}n de F{\'i}sica, Departamento de Astrof{\'i}sica}\\
\mbox{Avenida Astrof{\'i}sico Francisco S{\'a}nchez, s/n, Apartado 456, 38200 San Crist{\'o}bal de La Laguna (Tenerife), Espa\~na}
\vskip1pt
\mbox{\textsuperscript{h}Universit\`a degli Studi di Napoli, Federico II (UNINA), Dipartimento di Fisica Ettore Pancini}\\
\mbox{Via Cinthia 9, 80126 Napoli, Italia}
\vskip1pt
\mbox{\textsuperscript{i}Tomsk  State  University  of Control  Systems  and  Radioelectronics (TUSUR), Laboratory  for  Theoretical  Cosmology}
\mbox{40 prospect Lenina, 634050 Tomsk, Rossiya}
}

\date{25 November 2020}

\begin{abstract}
The total red shift $z$ might be recast as a combination of the expansion red shift and a static shift due to the 
energy-momentum tensor non-conservation of a photon propagating through Electro-Magnetic (EM) fields. 
If massive, the photon may be described by the de Broglie-Proca (dBP) theory which satisfies the Lorentz(-Poincar\'e) 
Symmetry (LoSy) but not gauge-invariance. The latter is regained in the Standard-Model Extension (SME), associated with 
LoSy Violation (LSV) that naturally dresses photons of a mass. The non-conservation stems from the vacuum expectation value of the vector and tensor LSV fields. 
The final colour (red or blue) and size of the static shift depend on the orientations and strength of the LSV and 
EM multiple fields encountered along the path of the photon. Turning to cosmology, for a zero $\Omega_{\Lambda}$ 
energy density, the discrepancy between luminosity and red shift distances of SNeIa disappears thanks to the recasting of $z$. 
Massive photons induce an effective dark energy acting `optically' but not dynamically. 
\end{abstract}

\keywords{Standard-Model Extension, Photons, Light Propagation, Cosmology, Supernovae}

\maketitle


\section {Introduction \label{intro}} 

Astronomy is almost entirely built up from information coming from Electro-Magnetic (EM) signals, interpreted with Maxwellian 
linear and massless electromagnetism. This latter is possibly an approximation of a broader theory, as Newtonian gravitation is of the Einsteinian one.
Observations  \cite{capozziello-prokopec-spallicci-2017,lopezcorredoira-2017,lopezcorredoira-2018} have led to the proposal that the 
universe contains up to 96\% dark matter and dark energy (both entities thus far remaining theoretically unexplained and experimentally undetected) and holds 
to general relativity, as the correct theory of gravitation. 
Others, unconvinced by these {\it ad hoc} ingredients, propose new theories of gravitation, though general relativity scores high marks in all tests so far.

Faced with this dichotomy and the respective pitfalls, we turn to the signals. Could a different interpretation of light 
lead to a {\it third option}? After all, modern physics - relativity and quantum mechanics - started with a new interpretation of light. Ultimately, 
physics is in any case obliged to examine its foundations through either the formulation of new types of matter and energy or embracing new 
conceptions of gravitation or  electromagnetism.

The photon is the only massless free particle of the successful Standard-Model (SM), lately {\it challenged} by the neutrino mass, 
the light mass of the Higgs boson and the absence of candidate particles for the dark universe. So let us focus on 
massive photons, leaving non-linear electromagnetism for future work. 

The official photon mass upper limit is $10^{-54}$ kg  \cite{tanabashietal2018}, but, as pointed out 
in \cite{retinospalliccivaivads2016}, such a limit \cite{Ryutov-1997,Ryutov-2007,Ryutov-2009} arises primarily from modelling rather than measurement. 

In 1922 de Broglie proposed a massive photon \cite{db22} and, through the group velocity dispersion of the latter, estimated the mass upper limit as $10^{-53}$ kg \cite{db23,debroglie-1924}. The origin of his conception of the 
photon mass resides in his theory of fusion, which couples two or more free Dirac fields to produce a scalar or vector Klein-Gordon 
equation, typically an electron-positron or neutrino-anti-neutrino duet
\cite{debroglie-1932a,debroglie-1932b,debroglie-1934a,debroglie-1934b,debroglie-1934c,debroglie-1934d,debroglie-1936}. In
\cite{debroglie-1936} the modified Maxwell equations were first written in a non-covariant form. Thus, de Broglie, and accidentally 
his disciple Proca \cite{proca-1936b,proca-1936c,proca-1936d,proca-1937}, laid down the first massive electromagnetism, compliant 
with the Lorentz(-Poincar\'e) Symmetry (LoSy), though not gauge-invariant. Contrary to common belief, the Proca Lagrangian describes 
generic composite particles, among which Proca states that the photons are massless `{\it doublets de masse nulle}' 
\cite{proca-1937}. For a comprehensive monograph see \cite{db40}.

Energy-momentum non-conservation already manifests itself in the Maxwellian theory for a photon crossing space-time dependent electromagnetic 
(EM) background fields (host galaxy, intergalactic and Milky Way). Instead, we show here that the de Broglie-Proca (dBP) massive 
photon undergoes energy-momentum non-conservation also in the case of constant EM fields, since it couples to the non-constant EM background potential. 

Motivated by the above {\it challenges}, the SM Extension (SME), based on the LoSy Violation (LSV), 
was put forward \cite{colladaykostelecky1997,colladaykostelecky1998}. 
The LSV breaks the invariance of the laws of physics for all positions and inertial velocities of the observers, and thus naturally  
accommodates anisotropies. The SM is LoSy invariant up to the energy scales at which the LSV occurred in the early universe. The SME 
allows the testing of the low-energy manifestations of the LSV in the present universe.  

The foundations of the analysis of light propagation in the SME were laid down in \cite{bodshnsp2017,bodshnsp2018} . Going beyond the SM, a 
  massive photon emerges by scrutiny of the group velocity dispersion or by recasting the SME Lagrangian.  
The effective, and frame-dependent, mass is proportional to the value of the LSV parameters and, in contrast to the dBP formalism, 
the SME massive photon is gauge-invariant \cite{bodshnsp2017,bodshnsp2018} and is compatible with the LSV upper limits 
\cite{kosteleckyrussell2011,gomesmalta2016}. Some of the benefits and drawbacks of attributing a mass to photons have been analysed recently 
\cite{craig-garciagarcia-2018,alfaro-soto-2019,reece-2019,govindarajan-kalyanapuram-2019,ferreira-etal-2020}.

Moreover, in specific conditions sub- and super-luminal velocities, imaginary and complex frequencies, and birefringence appear. Evidently in any condition, LSV anisotropy and inhomogeneity are present. 

Also in the SME, the photon energy-momentum tensor is not conserved in vacuum, and most remarkably, includes terms depending 
on constant LSV and EM fields \cite{bodshnsp2018}. Furthermore, we have determined 
 a non-expansion related frequency shift (towards the red or the blue) $z_{\rm LSV}$ \cite{helayel-spallicci-2019}. 

Quantified by the Hubble(-Humason)-Lema\^itre constant $H_0$  and detected  at the end of the '20s, the cosmological expansion 
stretches wavelengths from astrophysical sources, including Type Ia supernovae  (SNeIa). But since the end of the '90s, 
these latter, considering their faded luminosity with respect to the expected energy emitted, prompted the conception of dark 
energy, a mysterious fluid acting as the source of the cosmic accelerated expansion \cite{riess-etal-1998, perlmutter-etal-1999}.

If the cosmic flow expanded at a constant rate, an SNIa red shift would be directly proportional to its distance, and thus to its brightness.
Instead, the accelerating universe filled with dark energy implies that space expanded less quickly in the past than it does 
now. The light from distant objects such as SNeIa is less stretched during its journey to us, given how slowly space expanded during
 much of the time. An SNIa at a given distance (computed through its brightness, since SNeIa are believed to be standard candles) 
appears less red shifted than it would in a universe without dark energy. To put it another way, the red shift is not a measurement of
 distance but of expansion. When a distant source is observed, the lower expansion induces a lower red shift.  

The apparent acceleration of the younger Universe could be simply accounted for by a non-zero cosmological constant (originally 
proposed by Einstein to refute an expanding universe) on the left (geometrical) side of the field equations, or else by an extra 
term on the right (energy-matter) side acting as a source term for gravity, and corresponding to a vacuum energy density.

Various hypotheses were investigated concerning the nature of dark energy, {\it e.g.} \cite{peebles-ratra-2003,bamba-etal-2012}, 
among which was the extension of the gravity sector 
\cite{capozziello-francaviglia-2008,capozziello-delaurentis-2011,clifton-etal-2012,ezquiaga-zumalacarregui-2017}. 
 
Somewhat similarly, in this paper we also refer to an extension, namely that of the SM. Indeed, we identify the 
vacuum energy density in the framework of the SME as being due to the LSV field, which manifests itself through the $z_{\rm LSV}$ shift.

The above-mentioned {\it third option} may be checked by adding algebraically the $z_{\rm LSV}$ shift to the cosmological
 expansion shift $z_{\rm C}$ in the analysis of the data from SNeIa. We go through Baryonic Acoustic Oscillation (BAO), 
Cosmic Microwave Background (CMB) and gravitational lensing, often reputed to back the existence of dark energy, to see if 
any counter-evidence would threaten our hypothesis. We also deal with time dilation in SNeIa.

Attributing different origins to the nature of dark energy does not fully describe the scope of the debate 
that arose since the appearance of \cite{riess-etal-1998, perlmutter-etal-1999}. One school of criticism focuses on 
the reliability of SNeIa, BAO, CMB and gravitational lensing data, and challenges the counterarguments defending the 
evidence for dark energy based on data, {\it e.g.}, 
\cite{
vishwakarma-narlikar-2010,
nielsen-guffanti-sarkar-2016,
rubin-hayden-2016,
haridasu-lukovic-dagostino-vittorio-2017,
velten-gomes-busti-2018}. 

Another school of thought addresses alternatives to dark energy, of which we mention a few: 
photons oscillating into axions to produce a dimming effect for SNeIa \cite{csaki-kaloper-terning-2002a,csaki-kaloper-terning-2002b},
the different spectrum of primordial fluctuations generated by inflation \cite{blanchard-etal-2003},
timescape cosmology and the inhomogeneous Universe \cite{dam-heinesen-wiltshire-2017}, 
SNIa luminosity depending on red shift \cite{tutusaus-lamine-dupays-blanchard-2017,tutusaus-lamine-blanchard-2019},
stronger evidence of acceleration dipole approximately aligned with the CMB dipole than evidence of a monopole dark
energy acceleration \cite{colin-etal-2019,salehi-etal-2020}.  

Last but not least, the identification of SNeIa as standard candles is at risk of being breached, {\it e.g.}, \cite{morenoraya-etal-2016}. 

We wish to emphasise that we do not take a position on data reliability. Our main point is to recast the red shift, 
nowadays uniquely explained by expansion; our recasting may be tailored to whatever set of data on which consensus will be reached. Moreover, a lesser $z_{\rm LSV}$ shift value could coexist with dark energy. 

Generally speaking, the Concordance Lambda-Cold Dark Matter ($\Lambda$CDM) model may not rest on very firm ground, 
if Planck satellite CMB data, commonly reputedly its pillar, are reread and significantly support a closed universe \cite{divalentino-melchiorri-silk-2019}.  

In conclusion, our work is part of a large and scientifically healthy {\it afflatus} that revisits modern cosmology to verify its soundness.

Some wording on the relations between LSV, SME, extensions of the SM and photon mass are in order. LSV is not 
a sufficient condition to induce an effective mass. Generally speaking, extension of the SM such as the Minimally 
Supersymmetrised SM (MSSM), the Massive Neutrino Model (MNM), the Grand Unification Theory (GUT) do not induce an 
effective mass either. Thereby, it is the SME which is based on LSV that may cause an effective mass. Let us be more specific. 

The SME-LSV factors are represented by a $k^{\rm AF}_{\alpha}$ [metre $^{-1}$] 4-vector when the handedness of the 
Charge conjugation-Parity-Time reversal (CPT) symmetry is odd and by a $k_{\rm F}^{\alpha\nu\rho\sigma}$ [dimensionless] 
tensor when even. The $k^{\rm AF}_{\alpha}$ vector coming from the Carroll-Field-Jackiw Lagrangian \cite{cafija90} induces always a mass, while the $k_{\rm F}^{\alpha\nu\rho\sigma}$ tensor only in a supersymmetrised context after photino integration  \cite{bodshnsp2017,bodshnsp2018}.

The SME is built up by means of the Effective Field Theory approach and has the SM and General 
Relativity (GR) as possible limiting cases. In this paper, we work within a scenario such that the LSV does not leave its 
imprints in the GR sector. We instead focus on a special situation where the LSV is present in the photonic sector, 
through the $k^{\rm AF}_{\alpha}$ and the $k_{\rm F}^{\alpha\nu\rho\sigma}$ terms. The space-time metric, spin connection 
and curvature are unaffected by the LSV; we actually maintain Minkowski space-time with the anisotropies parametrised by 
$k^{\rm AF}$ and $k^{\rm F}$. This means that the effects of the LSV on the red shift stem exclusively from the presence
 of the violating parameters in the photonic sector. This is the viewpoint we adopt here. It is, however, a relevant matter, 
for future work to consider the presence of LSV in the gravitational sector 
\cite{
jacobson-mattingly-2003,
potting-2006,
kostelecky-russell-tasson-2008,
boldo-etal-2010,
blas-lim-2014,
eichhorn-platania-schiffer-2020} 
of SME and thereby reassess its effect on the red shift.

The rest of this paper is structured as follows. Section \ref{appa} shows the energy-momentum non-conservation 
of the dBP photon and prepares for the more difficult analysis in the SME dealt with in Sect.\ \ref{LightSME}, devoted to light 
propagation in this context; Sect.\ \ref{lsvde} focuses on the reinterpretation of dark energy through the $z_{\rm LSV}$ shift; 
Sections \ref{discussion} and \ref{perspectives} discuss the results and their implications. 

Our metric has a (+, -, -, -) signature; the Greek (Latin) indices run over 0,1,2,3 (1,2,3). We adopt SI units.

\section{Energy non-conservation in Maxwell and de Broglie-Proca theories\label{appa}}

We recall here a basic feature of the dBP theory concerning the 
non-conservation of the photon energy-momentum and pave the way to the analysis in the SME. We imagine the photon crossing an electromagnetic background. 

The dBP equations of motion derived from the Lagrangian \cite{proca-1936b} correspond to the divergence 
of the electric field and the curl of the magnetic field. They are given by 

\beq
\partial_\alpha F^{\alpha\beta}_{\rm T } + {\cal M}^2 A^\beta_{\rm T  } = \mu_0 j^\beta~,
\label{dBfe}
\eeq 
where ${\cal M} = {\ds \frac{m_\gamma c}{\hbar}}$, $m_\gamma$ is the photon mass, 
$c = 2.998 \times 10^8$ [m s$^{-1}$] is the speed of light and $\hbar = 1.055 \times 10^{-34}$ [kg m$^2$ s$^{-1}$].
 T stands for the total EM quantities due to background and photon contributions; $j^\beta$ is an external current, if it exists, 
and $\mu_0 = 1.257 \times 10^{-6}$ [kg m A$^{-2}$s$^{-2}$] is the magnetic permeability. 
Splitting the EM tensor field and the EM 4-potential in the background (capital letters) and photon (small letters) contributions, we have

\beq
A^\beta_{\rm T} = A^\beta + a^\beta ~~~~~~~~~~~~~~~~~F^{\alpha\beta}_{\rm T} = F^{\alpha\beta} + f^{\alpha\beta}~,
\eeq 
which substituted into Eq. (\ref{dBfe}) gives

\beq
\partial_\alpha f^{\alpha\beta} + {\cal M}^2 a^\beta = \mu_0 j^\beta - \partial_\alpha F^{\alpha\beta} - {\cal M}^2 A^\beta ~.
\label{dBfe-split}
\eeq 

Equation (\ref{dBfe-split}) tells us that the dBP photon interacts with the background 
through the potential even when the background field is constant. Indeed, if a field is constant, its associated potential is not

\beq
F^{\alpha\beta} = \partial^\alpha A^\beta - \partial^\beta A^\alpha~. 
\eeq

Conversely, this is not the case for the Maxwell photon, Eq. (\ref{M-split}), which interacts only with a non-constant field

\beq
\partial_\alpha f^{\alpha\beta} = \mu_0 j^\beta - \partial_\alpha F^{\alpha\beta}~.
\label{M-split}
\eeq 

The energy-momentum density tensor $\theta_{\ \tau}^{\alpha}$ [Jm$^{-3}$] for the dBP photon is obtained after a lengthy computation and is given by 

\begin{eqnarray}
\theta^\alpha_{~\tau} = 
& {\ds \frac{1}{\mu_0}} \left[ f^{\alpha\beta} f_{\beta\tau} + {\cal M}^2 a^\alpha a_\tau + {\cal M}^2 A^\alpha a_\tau + \right. \nonumber \\
& \left. \delta^\alpha_\tau \left( {\ds \frac{1}{4}} f^2 - {\ds \frac{1}{2}} {\cal M}^2 a^2 - {\cal M}^2 A^\beta a_\beta\right)\right]~.
\label{dBemt}
\end{eqnarray}

For $\vec E$, $\vec B$, the electric and magnetic fields of the background, $\vec e$, $\vec b$, the electric and magnetic fields of the photon, 
$\Phi$, $\vec A$, scalar and vector potentials of the background, and $\phi$, $\vec a$, scalar and vector potentials of the photon, 
we make explicit first the energy density [Jm$^{-3}$] 

\beq
\theta^0_{~0} =  \frac{1}{2}\left\{
\epsilon_0 e^2 
+ \frac{1}{\mu_0} b^2 
+ {\cal M}^2\left[\left(\epsilon_0\phi^2 + \frac{1}{\mu_0}a^2\right)
+ 2 \frac{{\vec A}\cdot{\vec a}}{\mu_0}\right]\right\}
\eeq
where $\epsilon_0 = 8.854 \times 10^{-12}$ [A$^2$s$^4$kg$^{-1}$m$^{-3}$] is the electric permittivity, 
and second we make explicit the generalised Poynting density vector [Jm$^{-3}$]    

\beq
\theta^i_{~0} = \frac{1}{\mu_0 c} \left[ \left. {\vec e}\times {\vec b}\right|_i 
+ {\cal M}^2\phi (a_i + A_i)\right]~.
\eeq

The energy-momentum density tensor  
variation $\partial_\alpha \theta_{\ \tau}^{\alpha}$ [Jm$^{-4}$] is given by 

\beq
\partial_\alpha \theta^\alpha_{~\tau} = 
\underbrace{
j^\alpha f_{\alpha\tau}- {\ds \frac{1}{\mu_0}}(\partial_\alpha F^{\alpha\beta})f_{\beta\tau}
}_{\text{Maxwellian terms}} 
\underbrace{ +
{\ds \frac{1}{\mu_0}} {\cal M}^2 (\partial_\tau A^\beta)a_\beta
}_{\text{de Broglie-Proca term}}~.
\label{dBnoncon}
\eeq 
which, after multiplication by $c$, in the explicit form becomes [Jm$^{-2}$s$^{-1}$]

\begin{align}
c & \left(\partial_0 \theta^0_{~0} + \partial_i \theta^i_{~0}\right) = 
\ \partial_t \theta^0_{~0} + c \ \partial_i \theta^i_{~0} = \nonumber\\
& - {\vec j}\cdot{\vec e} 
- \epsilon_0 \left (\partial_t {\vec E}\right)\cdot {\vec e} 
+ \frac{1}{\mu_0}\left(\nabla \times {\vec B}\right)\cdot{\vec e} ~ - \nonumber\\
& {\cal M}^2 \left[\epsilon_0\left(\partial_t \Phi\right)\phi - \frac{1}{\mu_0} \left (\partial_t{\vec A}\right)\cdot{\vec a}\right]~.
\end{align}

In conclusion, the energy-momentum density tensor of the dBP photon is not conserved. In addition to the 
Maxwellian terms, the mass couples with the background potential time-derivative. 

\section {Light propagation in the SME\label{LightSME}} 

\subsection {Non-conservation} 

In contrast to the LSV vector, the LSV tensor does not violate CPT. The frequency LSV 
shift that we shall be dealing with here is an observable of CPT violation, since it depends on the vector and tensor formulations.
Incidentally, we shall see that the leading term is $k^{\rm AF}_0$ which violates CPT.

Indicating by the symbol * the dual field, the photon energy-momentum density tensor $\theta_{\ \tau}^{\alpha}$ [Jm$^{-3}$] is
 \cite{bodshnsp2018,helayel-spallicci-2019}

\begin{align}
\theta_{\ \tau}^{\alpha} & = \frac{1}{\mu_0} \left( f^{\alpha\nu}f_{\nu\tau}
+ \frac{1}{4} \delta_{\tau}^{\alpha}f^{2}
- \frac{1}{2} k^{\rm AF}_{\tau}\ ^{*}f^{\alpha\nu}a_{\nu}+\right.\nonumber \\
 & \ \ \left. k_{\rm F}^{\alpha\nu\kappa\lambda}f_{\kappa\lambda}f_{\nu\tau}
+\frac{1}{4} \delta_{\tau}^{\alpha}k_{\rm F}^{\kappa\lambda\nu\beta}f_{\kappa\lambda}f_{\nu\beta}\right )~.
\label{pemt}
\end{align}

We render explicit the energy density [Jm$^{-3}$] 

\begin{align}
\theta^0_{~0} = &   
\frac{1}{2}\left[\epsilon_0 \left(\delta_{ij} - 2\chi_{ij}\right)e_i e_j + \right. \nonumber \\ 
& \left. \frac{1}{\mu_0} \left(\delta_{ij} + 2\zeta_{ij}\right)b_i b_j - k^{\rm AF}_{0}{\vec b}\cdot{\vec a}\right]~, 
\end{align}
where $k^{\rm AF}_0$ represents the time component of the breaking vector and the $k_{\rm F}^{\alpha\nu\kappa\lambda}$ tensor is 
decomposed as

\begin{align}
& k_{\rm F}^{0i0j} = \chi_{ij} = \chi_{ji}~,\\
& k_{\rm F}^{0ijk} = \epsilon_{ikl} \xi_{il}~,\\
& k_{\rm F}^{ijkl} = \epsilon_{ijm} \epsilon_{kln} \zeta_{mn}~; 
\end{align}
$\chi_{ij}$ and $\zeta_{m,n}$ are symmetric in the indexes, $\xi_{il}$ does not have symmetry properties; 
$\chi_{ij}$ and $\zeta_{m,n}$ have 6 components each, while $\xi_{il}$ has 9, totalling the 21 components of $k_{\rm F}^{\alpha\nu\kappa\lambda}$.   
We now render explicit the generalised Poynting density vector  [Jm$^{-3}$]    

\begin{eqnarray}
\theta^i_{~0} = 
& {\ds \frac{1}{\mu_0 c}} \left. {\vec e}\times {\vec b}\right|_i 
- 2 {\ds \frac{1}{\mu_0}} \left({\ds \frac{1}{c}}\epsilon_{ijl}\xi_{kl} e_k b_j + \epsilon_{ijk}\zeta_{kl} b_j b_l\right)- \nonumber \\
& {\ds \frac{1}{2}} k^{\rm AF}_0\left(\epsilon \phi e_i 
- {\ds \frac{1}{\mu_0 c}} \left. {\vec a}\times {\vec e}\right|_i \right)~.
& ~.
\end{eqnarray}

The energy-momentum density tensor variation $\partial_{\alpha} \theta_{\ \tau}^{\alpha}$ [Jm$^{-4}$] is given by 

\vspace*{-0.2cm} 

\begin{eqnarray}  
&\partial_{\alpha}\theta_{\ \tau}^{\alpha} = 
\underbrace{
j^{\nu}f_{\nu \tau} - {\ds \frac{1}{\mu_0}} \left(\partial_{\alpha}F^{\alpha \nu}\right)f_{\nu \tau}
}
_{\text{Maxwellian terms}} 
 \nonumber \\ 
& - {\ds \frac{1}{\mu_0}} \left [ 
\underbrace{ 
 \frac{1}{2}
\left(\partial_{\alpha}k^{\rm AF}_{\tau}\right)\ ^{*}f^{\alpha\nu}a_{\nu} - \frac{1}{4}\left(\partial_{\tau}k_{\rm F}^{\alpha \nu\kappa\lambda}\right)f_{\alpha \nu}f_{\kappa\lambda}
}
_{\text{EM background independent terms}} + 
\right. \nonumber \\
& 
\left.
\underbrace{\partial_{\alpha}\left(k_{\rm F}^{\alpha \nu\kappa\lambda}F_{\kappa\lambda}\right)f_{\nu \tau}
}
_{\text{non-constant term}} + 
\underbrace{
k^{\rm AF}_{\alpha}{^*}F^{\alpha \nu}f_{\nu \tau}
}
_{\text{constant term}}
\right]~. 
\label{pemt-nc-0}
\end{eqnarray}

After multiplication by $c$, the energy-momentum density tensor variation in the explicit form becomes [Jm$^{-2}$s$^{-1}$]

\begin{align}
~~~~ c & \left(\partial_0 \theta^0_{~0} + \partial_i \theta^i_{~0}\right) = 
 \partial_t \theta^0_{~0} + c \ \partial_i \theta^i_{~0} = \nonumber\\
& - {\vec j}\cdot{\vec e} 
- \epsilon_0 \left (\partial_t {\vec E}\right)\cdot {\vec e} 
+ \frac{1}{\mu_0}\left(\nabla \times {\vec B}\right)\cdot{\vec e} ~ - \nonumber\\
& \frac{1}{2} \frac{1}{\mu_0 c} \left\{ \left ( \partial_t k^{\rm AF}_{0}\right ) {\vec e} \cdot {\vec a} 
-  \left ( \nabla k^{\rm AF}_{0}\right ) \cdot \left [\phi{\vec e} + c \left ({\vec a} \times {\vec b}\right)\right ] \right\} +
 \nonumber \\
& \epsilon_0  \left ( \partial_t \chi_{ij}\right ) e_i e_j - 2 \frac{1}{\mu_0 c}  \left ( \partial_t \xi_{ij}\right ) e_i b_j 
+ 4 \frac{1}{\mu_0} \left ( \partial_t \zeta_{ij}\right ) b_i b_j + \nonumber \\
& 2 \epsilon_0  \partial_t \left (\chi_{ij}E_j\right ) \cdot e_i - 2 \frac{1}{\mu_0 c} \partial_t \left (  \xi_{ij}B_j\right ) \cdot e_i +
\nonumber \\ 
& \frac{1}{\mu_0 c} \left. {\vec e} \times \nabla \right|_i \cdot \left (2 \xi_{ji}E_j + c \zeta_{ij}B_j \right) - \nonumber \\
& \frac{1}{\mu_0 c} \left [ {\vec k}^{\rm AF} \cdot \left ( {\vec E} \times {\vec e}\right) + c k^{\rm AF}_{0} {\vec B}\cdot{\vec e}
\right]~.
\label{pemt-nc-0-expl}
 \end{align}

The energy-momentum density tensor variation is due to the following contributions: 
\begin{itemize}
\item{Maxwellian, LSV-independent terms that we have seen in Sect. \ref{appa}.}  \vskip5pt
There are three massive contributions (though not all the components are mass dependent).
\item{EM background independent terms implying that the energy-momentum density flux is not conserved in the absence 
of EM fields, if the LSV fields are space-time dependent. This is really a distinctive feature of the SME.}
\item{An LSV and EM space-time dependent term.} 
\item{A constant term coming solely from the CPT-odd handedness represented by the Carroll-Field-Jackiw (CFJ) 
electrodynamics \cite{cafija90}. Its action entails a non-constant 4-potential, for a constant EM background and a constant $k_{\rm AF}$. Indeed, there is an explicit $x^\alpha$ coordinate dependence at the level of the Lagrangian, exactly as in the dBP theory.}
\end{itemize}

The breaking tensor $k_{\rm F}$ appears either under a derivative or coupled to a derivative of the EM background.

{\renewcommand{\arraystretch}{1.2}
\begin{table}[h]
\centering
\caption
{\small 
{\footnotesize Upper limits of the LSV parameters, in SI units: 
$^{1}$Rotation in the polarization of light in resonant cavities \cite{gomesmalta2016}.  
$^{2}$Astrophysical observations \cite{kosteleckyrussell2011}. Such estimates are close to the Heisenberg limit 
on the smallest measurable energy or mass or length for a given time, $t$, set equal to the age of the universe.
 A full table of values was shown in \cite{helayel-spallicci-2019}.}}
\begin{tabular}{|c|c|}
\hline
$k^{\rm AF}_0$~~~~~~$^{1}$                               &            
$5.1 \times 10^{-10}$ m$^{-1}$         \\ \hline  
$k^{\rm AF}_0$~~~~~~$^{2}$                               &            
$5.1 \times 10^{-28}$ m$^{-1}$         \\ \hline  
\end{tabular}
\label{tableLSV}
\end{table}
}

The above contributions determine the energy-momentum non-conservation, according to the Noether theorem 
(symmetry breaking). Put another way, the photon exchanges energy with the LSV and EM fields. 

Comparing the non-conservation in dBP and SME cases, the former is relativistic and respects LoSy, while the 
latter is gauge-invariant. The advantage of the latter lies in tracing the origin of the mass in the LSV 
vacuum energy. In both cases a modification of the SM is necessary.  

\subsubsection{Origin of the non-conservation}

An estimate of the energy change that light would undergo was given \cite{helayel-spallicci-2019}. 
The wave-particle correspondence, even for a single photon \cite{aspect-grangier-1987}, leads us to consider 
that the light-wave energy non-conservation is translated into photon energy variation and thereby into a 
red or a blue shift. The energy variations, if there are losses, would translate into frequency damping.  

According to \cite{kosteleckysamuel1989b}, the LoSy breaking 4-vector, $k_{\rm AF}$, and the rank-4 tensor, 
$k_{\rm F}$, correspond, respectively, to the vacuum condensation of a vector and a tensor field in the 
context of string models. They describe part of the vacuum structure, which appears in the form of space-time
 anisotropies. Therefore, their presence on the right-hand side of Eq. 
(\ref{pemt-nc-0}) reveals that vacuum effects are responsible for the energy variation of light waves,
 which, in turn, correspond to a photon frequency shift. In plain terms, vacuum anisotropies are the really responsible 
for changing the frequency of light emitted by astrophysical structures.

\subsection{Sizing the LSV frequency shift}

In Eq. (\ref{pemt-nc-0}), we may neglect the tensorial perturbation in the frame of Super-Symmetry (SuSy) 
since it is less likely to condense than the CFJ vectorial perturbation
\cite{bebegahn2013,bebegahnle2015,helayel-spallicci-2019}. Independently of SuSy, the $k_{\rm F}$ term 
can be neglected since it is quadratic in the field strength and in frequency. The CFJ term instead contains a single derivative
and is linear in the frequency. Thus, for optical frequencies of the SNeIA, the $k_{\rm AF}$ 
is the dominant contribution. 

Excluding the space-time dependent components of the (inter-) galactic magnetic and LSV fields, only 
one contribution, the last term in Eqs. (\ref{pemt-nc-0},\ref{pemt-nc-0-expl}), survives. 
Finally, excluding external currents and large scale electric fields, an estimate was given along the line of sight observer-source for this term \cite{helayel-spallicci-2019},

\vspace*{-0.2cm} 
\beq
\partial_{\alpha}\theta_{\ 0}^{\alpha} \approx  
-\frac{1}{\mu_0} k^{\rm AF}_{0}\ ^{*}F^{0 i}f_{i 0}~, 
\label{pemt-nc-0-new-expl}
\eeq
where 
$^{*}F^{0 i}$ are the magnetic components of the dual EM background tensor field and 
$f_{i 0}$ are the electric components of the EM photon tensor field.

There is no relation between the $k^{\rm AF}_{0}$ and the photon mass\footnote{For simplicity of notation we replace 
$k^{\rm AF}$ by $V$; further, the 4-wave-vector is $k^{\mu}=\left({\ds \frac{\omega}{c}},\vec{k}\right)$, 
where $k^2 = \left({\ds \frac{\omega^2}{c^2}} - {\vec k}^2\right)$. Equations (3) in \cite{bodshnsp2017} 
or (9) in \cite{bodshnsp2018} confirms the CFJ dispersion relation \cite{cafija90}; that is,  
\begin{eqnarray}
&\left(k^{\mu}k_{\mu}\right)^{2}+\left(V^\mu V_\mu\right)\left(k^{\nu}k_{\nu}\right)-\left(V^\mu k_{\mu}\right)^{2}= \\
&\left[\left({\ds \frac{\omega}{c}}\right)^2 - {\vec k}^2 \right]^2 
+ \left(V_0^2 - {\vec V}^2\right)\left[\left({\ds \frac{\omega}{c}}\right)^2 - {\vec k}^2 \right]  
- \left(V_0{\ds \frac{\omega}{c}} - {\vec V}{\vec k} \right)^2= 0~. \nonumber
\label{CFJ DR}
\end{eqnarray}
For a massive photon, the rest mass is computed in the rest frame of the photon, $\vec k = 0$, rendering Eq. (\ref{CFJ DR}) as 
\beq
\left({\ds \frac{\omega}{c}}\right)^4 
+ \left(V_0^2 - {\vec V}^2\right) \left({\ds \frac{\omega}{c}}\right)^2  
- \left(V_0{\ds \frac{\omega}{c}} \right)^2= 0~.
\label{CFJ DR k=0}
\eeq
Equation (\ref{CFJ DR k=0}) has two solutions: one non-massive for $\omega=0$ (since gauge symmetry is not 
broken in CFJ formalism), the other for ${\omega = c |{\vec V}|}$. In conclusion, the time component of the 
CFJ perturbation vector does not contribute to the photon mass.}, though there are obviously massive terms in Eq. (\ref{pemt-nc-0}).  

Assuming that the energy variation of a light-wave corresponds to a frequency shift for a photon, we converted the energy variation given by Eq. (\ref{pemt-nc-0-new-expl}) into 
$\Delta\nu$, $\nu$ being the photon frequency. For $\nu = 486$ THz \cite{helayel-spallicci-2019}

\vspace*{-0.2cm} 

\beq
|\Delta\nu|_{\rm LSV}^{486~{\rm THz}} \approx 3.6 \times 10^{47} B~f_{i 0}~t_{\rm LB}~ k^{\rm AF}_0~{\varrho} ~,
\label{estimate3new}
\eeq
where $t_{\rm LB}$ is the look-back time and ${\varrho}$ is an arbitrary attenuation parameter that takes into 
account that the magnetic fields (host galaxy, intergalactic, Milky Way) estimated at $B = 5 \times 10^{-10} - 5 \times 10^{-9}$ T 
each, passed through by the photon, probably have different orientations and thus compensate for each other. 
We have not considered any relevant magnetic field at the source. 

For a numerical estimate of Eq. (\ref{estimate3new}), let us consider a source at $z=0.5$, $t_{\rm LB} = 1.57 \times 10^{17}$ s,
 $f_{i 0}= 3.79 \times 10^{-9}$ V~s~m$^{-2}$, an average $B = 2.75 \times 10^{-9}$ T. We wish to determine what the maximum level of $z_{\rm LSV}$ could be.

For $k^{\rm AF}_0$ the laboratory and astrophysical upper limits are $5.1 \times 10^{-10}$ m$^{-1}$ and $5.1 \times 10^{-28}$ m$^{-1}$, 
respectively \cite{kosteleckyrussell2011,gomesmalta2016}, see Table \ref{tableLSV}. 

This leads to a  value of the order of $10\%$ of the total $z$, for ${\varrho} \approx 8 \times 10^{-26}$ and 
${\varrho} \approx 8 \times 10^{-8}$, respectively. These estimates are {\it arbitrary} and just show that they can 
recover the largest frequency shifts in agreement with observational data from SNeIa.
Nevertheless, with different assumptions on $\varrho$ and $k^{\rm AF}_0$, we can get considerably smaller percentages of the rate 
$z_{\rm LSV}/z$ that are still in agreement with data.

Indeed, it is important to note that the {\it single} $z_{\rm LSV}$ shift from a {\it single} SNIa may be small or large, 
red or blue, depending on the amplitude and  orientations of the LSV (vector or tensor) and of the EM fields (host galaxy, 
intergalactic medium, Milky Way), and obviously the distance of the source. In any case, the final $z_{\rm LSV}$ is the result of 
accumulated shifts, both red and blue, encountered along the path. 

We have set upper limits starting from the physics; in the following, we shall set upper limits from the cosmological data in 
answer to the question of what range of values of $z_{\rm LSV}$ can be accommodated by cosmology. In following Sections, we 
shall refer to a generic $z_{\rm LSV}$ given by all the terms in Eq. (\ref{pemt-nc-0}).  

\subsection{Recasting $z$ \label{zmodels}} 

The existence of a photon frequency shift not due to the relative motion of the source and the observer 
belongs to the realm of physics. The role of such shifts in cosmology can attain three levels: replace 
totally or accompany to a certain degree the expansion shift. In the case of total replacement, we would 
return to the conception of a static universe, which still passes some tests but fails many others 
\cite{lopezcorredoira-2017,lopezcorredoira-2018}. At the other extreme, such a shift could be marginal 
for cosmology, but nevertheless be of relevance for fundamental physics. 

Here, we explore an intermediate option for which the static shift is superposed on the expansion shift. 

After recalling that the definition of $z = \Delta \nu/\nu_o$, where $\Delta \nu = \nu_{\rm e} - \nu_o$ is the difference between the observed $\nu_o$ and emitted 
$\nu_{\rm e}$ frequencies, or else $z = \Delta \lambda/\lambda_{\rm e}$ for the wavelengths, we pose the following conjecture: 
expansion causes $\lambda_{\rm e}$ to stretch to $\lambda_{\rm c}$; that is, $\lambda_{\rm c} = (1+z_{\rm C})\lambda_{\rm e}$. 

The wavelength $\lambda_{\rm c}$ could be further stretched or shrunk for the LSV shift to $\lambda_{\rm o} = 
(1+z_{\rm LSV})\lambda_{\rm c}  = (1+z_{\rm LSV}) (1+z_{\rm C}) \lambda_{\rm e}$. 
But since $\lambda_{\rm o} = (1+z)\lambda_{\rm e}$, we have 
$1 +  z = (1+z_{\rm C}) (1+z_{\rm LSV})$; thus
\vspace*{-0.2cm} 

\begin{eqnarray}
z = z_{\rm C} + z_{\rm LSV}  + z_{\rm C}z_{\rm LSV} \coloneqq z_{\rm o}~. 
\label{newz}
\end{eqnarray}
where $z_{\rm o}$ is the spectroscopically or photometrically observed $z$. The second order is non-negligible for larger $z_{\rm C}$. 

We model the behaviour of $z_{\rm LSV}$ with distance in three\footnote{We refrain form considering 
a fourth option proportional to the observed frequency and the distance.}  different ways, Table \ref{tabmodels}, 
according to whether the frequency variation is proportional to
\begin{itemize}
\item {the instantaneous frequency and to the distance,}
\item {the emitted frequency and the distance, or}
\item {just the distance.}
\end{itemize} 

The $z_{\rm C}$ shift stems from  the expansion of the universe, whereas the $z_{\rm LSV}$ would occur also along a static distance $r$. The $k_i$ parameters can take either positive (frequency increase) or negative (frequency decrease) values.  

\begin{widetext}

\begin{table}
\caption{\footnotesize LSV shift types. In the first column, the frequency variation is proportional to the 
instantaneous frequency and to the distance; in the second to the emitted frequency and the distance; 
in the third only to the distance; $k_{1,2}$ have the dimensions of Mpc$^{-1}$, $k_3$ of Mpc$^{-1}$s$^{-1}$. 
The positiveness of the distance $r$ constraints the shifts, see forth row.}
\label{tab1}
\newcolumntype{C}[1]{>{\centering\arraybackslash}m{#1}}
\setlength\extrarowheight{3pt}
\begin{tabular}
{ 
| C{1cm} 
| C{4cm} 
| C{4cm} 
| C{4cm} 
|
} 
\hline
Type
& ~~~ 1
& ~~~~~~~~~~~~ 2 
& ~~~~~~~~~~~~~ 3 
\\ \hline
$d\nu $  
& ~~~~ $k_1 \nu dr$
& ~~~~~~~~~~~~~ $k_2 \nu_{\rm e} dr$
& ~~~~~~~~~~~~~~ $k_3 dr$ 
\\ \hline
$\nu_{\rm o}$ 
& ~~~~~ $\nu_{\rm e}\exp^{k_1 r}$  
& ~~~~~~~~~~~~~~ $\nu_{\rm e}(1 + k_2 r)$         
& ~~~~~~~~~~~~~~ $\nu_{\rm e}+ k_3 r$ 
\\ \hline
$z_{\rm LSV}$ 
& $\exp^{- k_1 r} - 1$                      
& ~~~~~~~~~~~ $-{\ds \frac{k_2 r}{1 + k_2 r}}$  
& ~~~~~~~~~~~ $-{\ds \frac{k_3 r}{\nu_e + k_3 r}}$ 
\\[6pt] \hline
$r$           
& $-{\ds \frac{\ln(1+z_{\rm LSV})}{k_1}}$ 
& $~~~~~~-{\ds \frac {z_{\rm LSV}}{k_2(1 + z_{\rm LSV})}}$ 
& $~~~~~~~~~~~~~-{\ds \frac {\nu_{\rm e} z_{\rm LSV}}{k_3(1 + z_{\rm LSV})}}$ 
\\ \hline
$ r >0$ 
&\vtop{\hbox{\strut $z_{\rm LSV}>0~~{\rm for}~~ k_1 < 0$}\hbox{\strut $- 1 < z_{\rm LSV} < 0~~{\rm for}~~ k_1 > 0$}}          
&\vtop{\hbox{\strut $z_{\rm LSV}<-1~~{\rm or}~~z_{\rm LSV} >0~~{\rm for}~~k_2 < 0$}\hbox{\strut~$- 1 < z_{\rm LSV} < 0~~{\rm for}~~k_2 > 0$}}    
&\vtop{\hbox{\strut $z_{\rm LSV}<-1~~{\rm or}~~z_{\rm LSV} >0~~{\rm for}~~k_3 < 0$}\hbox{\strut~$- 1 < z_{\rm LSV} < 0~~{\rm for}~~k_3 > 0$}}   
\\ \hline
\end{tabular}
\label{tabmodels}
\end{table}

\end{widetext}

\section{The LSV frequency shift and dark energy \label{lsvde}}

\subsection{Supernovae: luminosity and red shift distances}

A greater than expected SNIa luminosity distance $d_{\rm L}$ for a given red shift led to the proposal of dark 
energy in order to reach consistency with the data \cite{riess-etal-1998, perlmutter-etal-1999}. We pursue the same consistency by using Eq. (\ref{newz}) instead.

We do not intend in this paper to propose a fully fledged alternative cosmology, but solely to take the first steps 
towards the {\it third option} presented in Section \ref{intro}. In the following we do not intend to state
strict conclusions for cosmology; instead, we limit ourselves to exploring whether existing data can accommodate our reinterpretation; 
namely, the recasting of $z$ through the additional non-expansion related shift. For this exploration, we pick a 
popular cosmology simulator \cite{wright-2006,wright-2018} and draw from it our initial considerations.    

The values that $z_{\rm C}$ should assume for a fixed $d_{\rm L}$ are computed  for two values (0.72 and 0) of 
$\Omega_\Lambda$ energy density and three values (0.7, 0.67, 0.74) of $h$ in Table \ref{mimicz}. We 
find that a small percentage correction of the red shift (from 0.01\% up to 10\%, a range we have proven feasible 
in Sect. \ref{LightSME}) allows us to recover the {\it same} luminosity distances.
 The explored range of $z$ values for SNeIa is (0-2). The most distant SNIa is at $z=1.914$ \cite{jones-etal-2013}. Inverting Eq. (\ref{newz}), we get 

\beq
z_{\rm C} = \frac{z - z_{\rm LSV}}{1 + z_{\rm LSV}}~.   
\label{newz-inv}
\eeq

If the $z_{\rm LSV}$ shift is blue, and thus negative, the photon gains energy, which implies that the 
cosmological $z_{\rm C}$ is higher than the observed $z$. If red, and therefore positive, the photon 
dissipates energy along its path, which implies that the cosmological $z_{\rm C}$ is smaller than the observed $z$. 

From a first glance at Table \ref{mimicz}, it appears that in most cases adding a negative blue 
$z_{\rm LSV}$ shift to the cosmological $z_{\rm C}$ is sufficient for agreement between $d_{\rm L}$ 
and $z$. Instead, the red shifts appear for $h=0.67$ and small values of $z$. 
It is also evident that for higher $h$ values a larger $z$ recasting is necessary, and that the shift is bluer. 
Patently, these are macroscopic and averaged indications. The next step necessitates picking each SNIa 
from a catalogue and evaluating the associated $z_{\rm LSV}$.  

Preliminary simulations show that $z_{\rm LSV}$ turns into red or blue shifts, depending also on other 
cosmological parameters, {\it e.g.}, the value of matter density $\Omega_{\rm m}$ or of the curvature 
$\Omega_{\rm k}$, and on the behaviour of $z_{\rm LSV}$ with frequency and distance, Table \ref{tabmodels}.

At the bottom of Table \ref{mimicz}, we have added two lines for $z = 4$ and $z =11$ to gauge the behaviour 
of $z_{\rm LSV}$ at large $z$: the most distant superluminous SN is at $z = 3.8893$ \cite{cooke-etal-2012}, 
and the most distant galaxy is at $z = 11.09$ \cite{oesch-etal-2016}.

The data in Table \ref{mimicz} can be commented on by momentarily adopting the usual terminology of accelerated 
expansion. The acceleration is not observable in our immediate neighbourhood; it would start to be 
noticeable further out and increase up to a maximum at $z \approx 0.5$. After this threshold value, 
the acceleration would decrease and for approximately $z \geq 4$, the so-called turning point, it would 
change sign corresponding to the deceleration phase associated with the older universe. 

How to reinterpret the above through a recast $z$? Far out, for approximately $z \geq 4$, $z_{\rm LSV}$ 
becomes definitely (and more and more) red. Since a red $z_{\rm LSV}$ shift corresponds to dissipation,
 we feel comforted by this easy interpretation for remote distances. 

Instead, the major difficulty is explaining why $z_{\rm LSV}$ starts from a small (blue or red) value in 
our immediate neighbourhood and increases up to a (blue) maximum for $z \approx 0.5$. This is maybe due 
to our location in a local void, or other manifestations of inhomogeneities or anisotropies. But before 
embarking in this sort of argument, the controversy on the data supporting dark energy \cite{
vishwakarma-narlikar-2010,
morenoraya-etal-2016,
nielsen-guffanti-sarkar-2016,
rubin-hayden-2016,
haridasu-lukovic-dagostino-vittorio-2017,
velten-gomes-busti-2018, 
csaki-kaloper-terning-2002a,
csaki-kaloper-terning-2002b,
blanchard-etal-2003,
dam-heinesen-wiltshire-2017, 
tutusaus-lamine-dupays-blanchard-2017,
tutusaus-lamine-blanchard-2019,
colin-etal-2019, 
kang-etal-2020,
salehi-etal-2020}
should be first of all settled. 

\begin{widetext}

\begin{table}[h]
\caption{\footnotesize We hold to the observed $z_{\rm o} \coloneqq z $, Eq. (\ref{newz}), and show the values 
that the cosmological shift $z_{\rm C}$ and possibly $z_{\rm LSV}$ should assume for obtaining a given luminosity distance $d_{\rm L}$, in the first column.  
We assume different values for $H_0 = h\times 100$ km/s per Mpc and for $\Omega_\Lambda$ densities (the curvature and radiation densities are set to zero, $\Omega_{\rm k} = \Omega_{\rm rad} = 0 $). In the second column for matter 
density $\Omega_{\rm m} = 0.28$, energy density $\Omega_{\Lambda} = 0.72$ and $h=0.7$, we pose $z_{\rm LSV} = 0$ 
and thereby $ z = z_{\rm C}$; in the third column for $\Omega_{\rm m} = 0.28$ but $\Omega_{\Lambda} = 0 $ 
and $h = 0.7$, we show the values of $z_{\rm C}$ that determine the same $d_{\rm L}$; in the fourth, eighth and 
twelfth columns, the percentage variation of $z_{\rm C}$; in the fifth,  ninth and thirteenth columns, from 
Eq. (\ref{newz}), $z_{\rm LSV} = {\ds \frac{z - z_{\rm C}}{1 + z_{\rm C}}}$; in the sixth, tenth and 
fourteenth  columns, the rate ${\ds \frac{z_{\rm LSV}}{z}} $; 
in the seventh column for $\Omega_{\rm m} = 0.28$ but $\Omega_{\Lambda} = 0 $ and $h = 0.67$, the values 
of $z_{\rm C}$ which determine the same $d_{\rm L}$; in the eleventh column for $\Omega_{\rm m} = 0.28$ 
but $\Omega_{\Lambda} = 0 $ and $h = 0.74$, the values of $z_{\rm C}$ which determine the same $d_{\rm L}$. 
Red  or 
blue shifts correspond to positive and negative values of $z_{\rm LSV}$, respectively. 
The most distant SNIa is at $z=1.914$ \cite{jones-etal-2013}, the most distant superluminous SN is at 
$z = 3.8893$ \cite{cooke-etal-2012}, and the most distant galaxy is at $z = 11.09$ \cite{oesch-etal-2016}. 
The numerical values are derived from a Cosmology Simulator \cite{wright-2006,wright-2018}.}
\label{tab1}
\newcolumntype{C}[1]{>{\centering\arraybackslash}m{#1}}
\setlength\extrarowheight{3pt}
\begin{tabular}
{ 
| C{1.19cm} 
| C{1.48cm} 
|| C{1.48cm} 
| C{1.0cm} 
| C{1.3cm} 
| C{0.9cm} 
|| C{1.48cm} 
| C{1.0cm} 
| C{1.3cm} 
| C{0.9cm} 
|| C{1.48cm} 
| C{1.0cm} 
| C{1.3cm} 
| C{0.9cm} 
|
} 
\hline
I 
& II 
& III
& IV
& V
& VI
& VII
& VIII
& IX
& X
& XI
& XII
& XIII
& XIV
\\ \hline
$ d_{\rm L}$ [Gpc]
& $h = 0.7$ $\Omega_{\rm m} = 0.28$ $\Omega_{\Lambda} = 0.72$ $z_{\rm LSV} = 0$ $ z = z_{\rm C}$ ------------ $z$ 
& $h = 0.7$ $\Omega_{\rm m} = 0.28$ $\Omega_{\Lambda} = 0$ $z_{\rm LSV} \neq 0$, $z \neq z_{\rm C}$ ----------- $ z_{\rm C}$ 
& ${\ds\frac{z_{\rm C}- z}{z}} $ [\%]
& $z_{\rm LSV}$
& ${\ds \frac{z_{\rm LSV}}{z}}$  [\%]
& $h = 0.67$ $\Omega_{\rm m} = 0.28$ $\Omega_{\Lambda} = 0$ $z_{\rm LSV} \neq 0$, $z \neq z_{\rm C}$ ------------ $ z_{\rm C}$  
& ${\ds\frac{z_{\rm C}- z}{z}} $ [\%]
& $z_{\rm LSV}$
& ${\ds \frac{z_{\rm LSV}}{z}}$ [\%]
& $h = 0.74$ $\Omega_{\rm m} = 0.28$ $\Omega_{\Lambda} = 0$ $z_{\rm LSV} \neq 0$, $z \neq z_{\rm C}$ ----------- $ z_{\rm C}$  
& ${\ds\frac{z_{\rm C}- z}{z}} $ [\%]
& $z_{\rm LSV}$
& ${\ds \frac{z_{\rm LSV}}{z}}$ [\%]
\\ \hline
  $ ~~~0.2225 $  
& $ ~0.05000 $  
& $ ~~0.05063 $ 
& $ ~~~~1.26 $ 
& $ -0.00059 $ 
& $ -1.19 $
& $ ~~0.04872 $
& $ ~-2.56 $
& $ ~~~0.00122 $
& $ ~~2.44 $
& $ ~~0.05368 $
& $ ~~~~7.36 $
& $ -0.00349 $
& $ -6.98 $
\\ \hline
  $ ~~0.4610 $  
& $ ~0.10000$  
& $ ~~0.10314$ 
& $ ~~~~3.14$ 
& $ -0.00285 $ 
& $ -2.85 $
& $ ~~0.09888 $
& $ ~-1.12 $
& $ ~~~0.00107 $
& $ ~~1.07 $
& $ ~~0.10877 $
& $ ~~~~8.77 $
& $ -0.00791 $
& $ -7.91 $
\\ \hline
  $ ~~2.8528 $  
& $ ~0.50000$  
& $ ~~0.54649$ 
& $ ~~~~9.30$ 
& $ -0.04425 $
& $ -8.88 $
& $ ~~0.52645 $
& $ ~~~~5.29 $
& $ -0.01733 $
& $ -3.46 $
& $ ~~0.57322 $
& $ ~~14.64 $
& $ -0.04654 $
& $ -9.31 $
\\ \hline
  $ ~~6.6874 $ 
& $ ~1.00000$  
& $ ~~1.10489$ 
& $ ~~10.49 $ 
& $ -0.04983$        
& $ -4.98 $
& $ ~~1.06682 $
& $ ~~~~6.68 $
& $ -0.03233 $
& $ -3.23 $
& $ ~~1.15473 $
& $ ~~15.47 $
& $ -0.07181 $
& $ -7.81 $
\\ \hline
  $ ~11.0776 $ 
& $ ~1.50000$  
& $ ~~1.63897$ 
& $ ~~~9.26 $
& $ -0.05266$         
& $ -3.51 $
& $ ~~1.58473 $
& $ ~~~~5.65 $
& $ -0.03278 $
& $ -2.18 $
& $ ~~1.71022 $ 
& $ ~~14.01 $ 
& $ -0.07756 $
& $ -5.17 $
\\ \hline
  $ ~15.8128$ 
& $ ~2.00000$  
& $ ~~2.14731$  
& $ ~~~7.36 $
& $ -0.04681$  
& $ -2.34 $
& $ ~~2.07771 $
& $ ~~~~3.88 $
& $ -0.02525 $
& $ -1.26 $
& $ ~~2.23873 $
& $ ~~11.93 $
& $ -0.07371 $
& $ -3.68 $
\\ \hline
\hline 
  $ ~36.6276$ 
& $ ~4.00000$  
& $ ~~3.99729$  
& $ ~-0.07 $
& $ ~~0.00054$  
& $ ~~~0.01 $
& $ ~~3.87115 $
& $ ~-3.22 $
& $ ~~0.02645 $
& $ ~~0.66 $
& $ ~~4.16317 $
& $ ~~~~4.08 $
& $ -0.03160 $
& $ -0.79 $ 
\\ \hline
  $ 118.5408$ 
& $ 11.00000$  
& $ ~~9.47515$  
& $ -13.86 $
& $ ~~0.14557$  
& $ ~~~1.32 $
& $ ~~9.17182 $
& $ -16.62 $
& $ ~~0.17973 $
& $ ~~1.63 $
& $ ~~9.87518 $
& $ -10.22 $
& $ ~~0.10343 $
& $ ~~0.94 $ 
\\ \hline
\end{tabular}
\label{mimicz}
\end{table}

\end{widetext}

\subsubsection{Time dilation and Supernovae \label{TDSN}}

Since the SME induces an effective mass to the photon \cite{bodshnsp2017,bodshnsp2018}, we are dealing with massive photons.
Blondin {\it et al.} \cite{blondin-etal-2008} compared several spectra from the more distant SNeIa with those 
of nearer ones and found that the more distant explosions took longer to unfold. They pointed out that time dilation in SNeIa cannot be attributed to `tired light'. 
Let us discuss not only this statement in the context of this paper, but also our assumptions on reinterpreting $z$ by examining three related issues.  

Firstly, just attributing 
mass to a photon does not determine {\it per se} a decaying frequency, a feature of the `tired light'.    
Indeed, massive photon propagation obeys the dBP equation which, in the absence of an EM background, is written as \cite{db40}

\beq
\left[\square + \left(\frac{m_\gamma c}{\hbar}\right)^2\right]a^\alpha = 0~, 
\label{dBPwave}
\eeq
$m_\gamma$ being the photon mass, $c$ the speed of light, and $\hbar$ the reduced Planck constant. Equation 
(\ref{dBPwave}) does not entail frequency decay, unless the mass has an imaginary component and there is a singular source term on the right-hand side
\cite{thiounn-1960,yourgrau-woodward-1974}. 
For violating photon energy conservation, apart an external current term, there must be an EM background field,  Eq. (\ref{dBnoncon}). 

Secondly, the energy variation computed here occurs regardlessly of the expansion. The associated shift has 
been named $z_{\rm LSV}$ and evaluated to be at most 10\% of the total $z$. It is legitimate to ask 
whether there is any impact on time dilation due to this reinterpretation of $z$.

In an expanding universe, the ratio of the observed frequency to that emitted by a distant object or the 
observed rate of any time variation in the intensity of the emitted radiation will be proportional to  
\beq
\frac{1}{1 + z}~. 
\eeq

Blondin {\it et al.} \cite{blondin-etal-2008} estimated $(1+z)^b$ as $b=0.97 \pm 0.10$, $b =1$ being  the value 
predicted by the Friedmann-Lema\^{i}tre-Robertson-Walker cosmology. A variation of 10\% for $b$ is tantamount to 
a variation of the same order for $z$, as the Taylor expansion shows 

\beq
\frac{1}{({1 + z})^b}\simeq 1 - bz + \frac{1}{2} (b+b^2) z^2~. 
\eeq
 
Therefore, supposing the total $z$ as the sum of an expansion related $z_{\rm C}$ and  a static $z_{\rm LSV}$ is 
compatible with the findings in \cite{blondin-etal-2008}. 
  
Thirdly, the massive photon group velocity differs from $c$ by a quantity proportional to the inverse of the frequency squared 
\cite{db22,db23,debroglie-1924,db40}. 
Such a dependence has been analysed recently with the signals from Fast Radio Bursts (FRBs) 
\cite{boelmasasgsp2016,wuetal2016b,boelmasasgsp2017,bebosp2017,shao-zang-2017,wei-zhang-zhang-wu-2017,wei-xu-2018,yang-zhang-2017,xing-etal-2019,wei-xu-2020}.

It appears that the SNIa spectra often shift from higher, $f_{\rm h}\approx 8 \times 10^{14}$ Hz, to lower, 
$f_{\rm l}\approx 4 \times 10^{14}$ Hz, optical frequencies during the burst \cite{blondin-etal-2008}. 
Photons emitted at the end will therefore take more time to reach the observer than those at start. 
This delay would occur even when the source is static, mimicking time dilation. For a source at distance $d$, 
the difference in arrival times is to be added to the burst duration. In SI units, the difference in arrival times is \cite{bebosp2017} 

\beq
\Delta t = \frac{d c^3m_\gamma^2}{8 \hbar^2\pi^2 }
\left(\frac{1}{f_{\rm l}^2} - \frac{1}{f_{\rm h}^2}\right)
\simeq  \frac{d}{c} 
\left(\frac{1}{f_{\rm l}^2} - \frac{1}{f_{\rm h}^2}\right) 
10^{100} m_\gamma^2~.
\label{deltat}
\eeq

Inserting the official upper limit on photon mass $10^{-54}$ kg, $d = 1.4\times 10^{9}$ light-years 
(approximately corresponding to $z=0.1$), $\Delta t$ is in the order of 
$10^{-21}$ s. In an expanding universe, Eq. (\ref{deltat}) becomes \cite{boelmasasgsp2016,boelmasasgsp2017}

\beq
\Delta t = \frac{1}{H} 
\left(\frac{1}{f_{\rm l}^2} - \frac{1}{f_{\rm h}^2}\right) 
10^{100} m_\gamma^2 H_\gamma~, 
\label{deltatcosm}
\eeq 
where 
\beq
H_\gamma = \int_0^z \frac{(1+z')^{-2}}{\sqrt{\Omega_m(1+z')^3 + \Omega_\Lambda}}dz'
\label{hgamma}
\eeq

Equation (\ref{hgamma}) reduces the outcome of Eq. (\ref{deltat}) by a factor 10 for common values of 
the $\Omega$ densities. Hence, we have shown that although massive photons determine an effect similar 
to time dilation for not-moving sources or add a static contribution to moving sources, such an effect is utterly  marginal for SNeIa. 

In conclusion, our assumptions and results are compatible with current literature on SNeIa.  

\subsection{BAO, CMB and gravitational lensing}

In Sect. \ref{intro}, we gave a brief account on the debates on i) the nature of dark energy, ii) the data 
reliability from SNeIa, BAO, CMB and gravitational lensing, and iii) the alternatives to dark energy. The latter 
two debates are not dominant as there is a prevailing tendency to converge towards the $\Lambda$CDM Concordance Model. 

Nevertheless, there is an increasing interest in discussing the foundations of common assumptions. A critical 
analysis of the supposed confirmation of dark energy by the Baryonic Atomic Oscillations (BAO) and by the 
Cosmic Microwave Background (CMB) has recently appeared \cite{tutusaus-lamine-dupays-blanchard-2017,tutusaus-lamine-blanchard-2019} in which it is shown that 
that a non-accelerated universe is nicely  able to fit low and high red shift data from SNeIa, BAO and 
the CMB on the main assumption of SN luminosity dependency on red shift. Now, Kang {\it et al.} \cite{kang-etal-2020} 
have found a significant (a 99.5\%) correlation between SN luminosity and stellar population age. It is worth 
recalling that the CMB can be interpreted differently, possibly without dark energy \cite{lopezcorredoira-2013} 
and in any case outside a canonical flat universe \cite{divalentino-melchiorri-silk-2019}. 

We have interpreted $z_{\rm LSV}$ as due to vacuum energy, an effective dark energy of LSV origin, although not 
producing acceleration of the expansion. We do not dispute the results from BAO and the CMB proving the 
existence of dark energy {\it per se}. Our interpretation would be falsified, if there were a strict proof of acceleration (m/s$^2$) with respect to an inertial frame.      

Further, it is worth considering that BAO and CMB data can accommodate our recasting of $z$ due to the 
uncertainties of spectroscopic and especially photometric measurements. In particular for the CMB, the photons 
were all emitted at the same time and same distance from us. Thus, any change in the definition of the red shift should not bear huge consequences. 

For BAO, the angular size of the sound horizon is measured for the large-scale structure at different red shifts, 
which allows us to measure $H(z)$, which depends on $\Omega_{\rm m}$, $\Omega_\Lambda$ and other variables.
Since $d_{\rm A}$ (angular distance) is proportional to $d_{\rm L}/(1+z)^2$ and the angular size of an object 
 is inversely proportional to $d_{\rm A}$, it would not be surprising that the same $z_{\rm LSV}$ mimicking the dark energy for SNeIa, acts similarly for the BAO peak. 

Finally, we tackle gravitational lensing. The classical method used to probe dark energy is to measure the ratio of average tangential shear 
as a function of the red shift behind the clusters \cite{heavens-2009}. This ratio depends on the relation between distance and red shift,
which is modified by dark energy, or by $z_{\rm LSV}$.

\section{Discussion of the main results\label{discussion}}

We have assumed the total red shift $z$ as a combination of the expansion red shift $z_{\rm C}$ and of a static, red or blue shift
$z_{\rm LSV}(r)$, $r$ being the travelled distance. The latter shift is due to the energy non-conservation of a 
photon propagating through EM background fields (host galaxy, intergalactic and Milky Way). 
Beyond the Maxwellian contributions, if the photon is massive, such propagation may be described within the framework 
of de Broglie-Proca (dBP) formalism or within the framework of the Lorentz(-Poincar\'e) Symmetry Violation (LSV) associated 
with the Standard-Model Extension (SME). In the latter case, the non-conservation stems from the vacuum 
expectation value of the vector and tensor LSV fields. 

Our understanding is that $z_{\rm LSV}$ is a manifestation of an effective dark energy caused by the expectation 
values of the vacuum under LSV. Indeed, we suggest that dark energy, {\it i.e.} LSV vacuum energy, is not causing 
an accelerated expansion but a frequency shift. 

The {\it single} $z_{\rm LSV}$ shift from a {\it single} SNIa may be small or large, red or blue, depending on the 
orientations of the LSV (vector or tensor) and of the EM fields (host galaxy, intergalactic medium, Milky Way), as 
well as the distance of the source. In any case, the colour of $z_{\rm LSV}$ is the final output of a series of shifts, both red and blue, encountered along the path. 

If the $z_{\rm LSV}$ shift is blue, thus negative, the photon gains energy; it implies that the real red shift  
is larger than the measured $z$. If red, thus positive, $z_{\rm LSV}$ corresponds to dissipation along the photon path; it implies that the real red shift is smaller than the measured $z$.

Recasting $z$, on average, we observe a blue static shift for $z \leq 2$, but a red one in our local Universe for 
smaller values of the Hubble(-Humason)-Lem\^aitre parameter ($67-74$ km/s per Mpc), and always red for $z > 4$.

The peculiarity of our approach is that a single mechanism could explain all the positions of the SNeIa in the 
($\mu, z$) plan, $\mu$ being the distance modulus, including the outliers. The experimental and observational 
limits on LSV and magnetic fields are fully compatible with our findings.   

\section{Perspectives \label{perspectives}}

The LSV shift provides a physical explanation of red shift remapping \cite{bassett-etal-2013,wojtak-prada-2016,wojtak-prada-2017,tian-2017}
and is not limited to the SNIa case. It is naturally suited to explaining recently discovered expansion anisotropies 
\cite{morenoraya-etal-2016,colin-etal-2019,migkas-etal-2020,salehi-etal-2020}. 

In future work, leaving aside massive photons, we will show that a frequency shift is also produced by a generalised 
non-linear electromagnetism, encompassing the formulations of Born and Infeld, and Euler, (Kockel) and Heisenberg. The calculated limits on 
$z_{\rm LSV}$ will be applicable to $z_{\rm NL}$ for the non-linear electromagnetism. 

Apart from an additional shift occurring also in Maxwellian electromagnetism in the presence of a space-time dependent EM 
field, we will show that departing from Maxwellian electromagnetism in three different directions (classically massive 
dBP electromagnetism or non-linear electromagnetism or the SME) leads to a common 
conclusion: the red shift is not only due to expansion. If and by how much such a departure is relevant for cosmology is the real question.  

Nineteenth century Maxwellian electromagnetism and the more modern Einsteinian gravitation have been well tested. This 
has not impeded the proposition of alternative formulations of electromagnetism and  gravity. The lack of 
experimental proof on the dark universe and the successes of general relativity prompt us to revisit astrophysical 
observations, largely based on light signals, with non-Maxwellian electromagnetism, opening the door to radically new interpretations.  

In future explorations, we will have to deal with the comparison of $z_{\rm LSV}$
with the error on $z$. The analysis of the error on spectroscopic and photometric measurements is destined to become a pivotal issue for cosmology 
\cite{palanquedelabrouille-etal-2010,calcino-davis-2017,davis-hinton-howlett-calcino-2019}. 

The $z_{\rm LSV}$ shift considered here is {\it at most} 10\% of the $z_{\rm C}$ expansion shift relative to $H_0 = 70$ km/s per Mpc and is 
thus below $2.3 \times 10^{-19} \Delta \nu/\nu$ per m (units for a static red shift). It would be desirable to test frequency 
invariance in vacuo \cite{shamirfox1967,wolf1986,wolf1987a,wolf1987b} with a ground or space based interferometer.  

\section*{Acknowledgments}

Discussions with M. A. Lopes Capri, V. E. Rodino Lemes, R. de Oliveira Santos (Rio de Janeiro), M. Benetti and F. Ragosta (Napoli) 
are acknowledged. SC acknowledges the support of INFN, {\it iniziative specifiche} QGSKY and MOONLIGHT2. The authors thank P. Brockill (University of Wisconsin-Milwaukee) for comments and T. Mahoney (La Laguna) for editing support.

\bibliographystyle{apsrev} 


\end{document}